\documentclass[aps,pra,twocolumn,showpacs]{revtex4-1}
\usepackage{graphicx}
\usepackage{amsmath,amsfonts}
\usepackage[T1]{fontenc}
\usepackage{hyperref}
\newcommand{\bra}[1]{%
  \left| #1 \right\rangle%
}

\newcommand{\me}[1]{%
  \langle #1 \rangle%
}
\newcommand{\var}[1]{%
  \langle \Delta {#1}^{2} \rangle%
}
\begin{document}
% title and author
\title{Quantum dynamics generated by the two-axis counter-twisting Hamiltonian}
\author{Dariusz Kajtoch and Emilia Witkowska}
\affiliation{Institute of Physics PAS, Aleja Lotnik\'ow 32/46, 02-668 Warszawa, Poland}
% abstract
\begin{abstract}
We study the quantum dynamics generated by the two-axis counter-twisting Hamiltonian from an initial spin coherent state in a spin-$1/2$ ensemble.
A characteristic feature of the two-axis counter-twisting Hamiltonian is the existence of four neutrally stable and two saddle unstable fixed points.
The presence of the last one is responsible for a high level of squeezing.
The squeezing is accompanied by the appearance of several quantum states of interest in quantum metrology with Heisenberg-limited sensitivity, and we show fidelity functions for some of them.
We present exact results for the quantum Fisher information and the squeezing parameter.
Although, the overall time evolution of both changes strongly with the number of particles, we find that they have regular dynamics for short times. 
We explain scaling with the system size by using a Gaussian approach.
\end{abstract}

\maketitle
% ----------------------------------------
% Section 1: Introduction
% ----------------------------------------
\section{Introduction}

Kitagawa and Ueda in their pioneering work~\cite{Kitagawa1993} have proposed the one-axis twisting (OAT) and two-axis counter-twisting (TACT) Hamiltonians for dynamical generation of spin-squeezed states. 
The Bose-Einstein condensate of ultra-cold atoms offers an exceptional tool for experimental realization of the one-axis twisting Hamiltonian and investigation of the many-particle entanglement. 
Several experiments have reported the creation of spin-squeezed states, by manipulation of internal states of multicomponent condensates~\cite{Riedel2010,Gross2010,Hamley2012},
or alternatively in a single-component condensate in a double-well potential~\cite{Esteve2010,Maussang2010}. 
In contrary, the two-axis counter-twisting Hamiltonian cannot be simply realized by inter-atomic interactions among ultra-cold atoms. 
Therefore several schemes were proposed to transform the one-axis twisting Hamiltonian into an effective two-axis counter-twisting Hamiltonian \cite{prop_TACT_doublewell_2003,OAT_to_TACT2011,OAT_to_TACT_Shen2013,OAT_to_TACT_suzuki_2014, Opatrny} or implement TACT model in other realistic systems~\cite{prop_TACT_2001, AndreLukin,prop_TACT_2002, prop_TACT_2003}. Actually, there is growing interest in quantum states generated by the TACT Hamiltonian since they support Heisenberg-like sensitivity of high-precision measurements and a much higher level of squeezing that is unachievable by OAT interactions~\cite{Kitagawa1993, Yukawa2014}.

In this paper, we study in great detail the quantum dynamics generated by the two-axis counter-twisting Hamiltonian from an initial spin coherent state in a spin-$1/2$ ensemble.
We start with the mean-field description which identifies a convenient location of the initial coherent spin state.
A characteristic feature of the two-axis counter-twisting Hamiltonian is the existence of the four neutrally stable and two saddle unstable fixed points.
The presence of the last one is responsible for high level of squeezing. 
On the other hand, the quantum dynamics around a stable fixed point generates states that support shot-noise limited sensitivity for quantum metrology.

The location of the initial spin coherent state on an unstable fixed point leads to strong stretching of the state along a meridian resulting in highly reduced variance of the spin operator. 
It has to be noted that the angle between the inflowing and outflowing trajectories in the phase space is $\pi/2$, thus the squeezing is optimal.
The high level of squeezing is accompanied by the presence of several quantum states of interest in quantum metrology with Heisenberg-limited sensitivity. We will show high fidelity to the Berry-Wiseman, 
equally-weighted superposition and Yurke states, and a bit worse to the twin-Fock state, as it was partially reported in \cite{Yukawa2014}.
We show that better sensitivity can be reached by using states produced by the dynamics at later times than the time for the optimal squeezing.

We calculate the quantum Fisher information in order to quantify the amount of quantum correlations generated in time that are useful for precision measurements.
Although, the overall time evolution of the quantum Fisher information changes strongly when one changes the number of particles $N$, 
we find that it has regular dynamics for short times of interest. 
In this regime, the time scales like $\sim\ln (2\pi N)/N$, and we explain the scaling using a Gaussian approach within the Bogoliubov-Born-Green-Kirkwood-Yvon hierarchy~\cite{AnglinVardi, AndreLukin}, 
as well as the known scaling of the variance of the spin operator.
%In addition, we calculate Wigner functions that visualize quntum states on the Bloch sphere, and also visualize the substantial multi-particle entanglement having fringes of negative values.
Our results show that the quantum dynamics with the two-axis counter-twisting Hamiltonian creates quantum correlations in a regular way on a short time scale. 
This time scale is reduced with an increased number of particles.

% ----------------------------------------
% Section 2: Model hamiltonian
% ----------------------------------------
\section{The model}

We consider a collection of $N$ qubits e.q. particles in two modes.
The system is conveniently described using the collective spin operator $\hat{\vec{S}}$,
whose components written in the Schwinger representation are
  \begin{subequations}
    \begin{align}
      \hat{S}_{x} &= \frac{1}{2}\left(\hat{a}^{\dagger}\hat{b} + \hat{b}^{\dagger}\hat{a} \right),\\
      \hat{S}_{y} &= \frac{1}{2i}\left(\hat{a}^{\dagger}\hat{b} - \hat{b}^{\dagger}\hat{a} \right),\\
      \hat{S}_{z} &= \frac{1}{2}\left(\hat{a}^{\dagger}\hat{a} - \hat{b}^{\dagger}\hat{b} \right),
    \end{align}
  \end{subequations}
where $\hat{a}^{\dagger}$, $\hat{b}^{\dagger}$ are creation operators associated with two modes.
The two-axis counter-twisting Hamiltonian proposed by Kitagawa and Ueda \cite{Kitagawa1993} is
  \begin{equation}\label{eq:two_axis_hamiltonian}
  \hat{\mathcal{H}}_{_{\rm TACT}} = \frac{\hbar \chi}{2i}\left(\hat{S}_{+}^{2} - \hat{S}_{-}^{2}\right),
  \end{equation}
where $\hat{S}_{\pm}=\hat{S}_x\pm i \hat{S}_y$. Nevertheless, we will operate on the rotated Hamiltonian
$\hat{\mathcal{H}} = \hat{U} \hat{\mathcal{H}}_{_{\rm TACT}} \hat{U}^{\dagger}$ with 
$\hat{U} = e^{-i (\pi/2) \hat{S}_{y}}$,
\begin{equation}\label{eq:two_axis_hamiltonian_exact}
\hat{\mathcal{H}} = - \hbar\chi\left(\hat{S}_{y}\hat{S}_{z} + \hat{S}_{z}\hat{S}_{y} \right)\text{.}
\end{equation}
Although, the action of a $SU(2)$ group element on (\ref{eq:two_axis_hamiltonian}) does not change the overall result, it describes a distinct physical system.
In our case, the Hamiltonian (\ref{eq:two_axis_hamiltonian_exact}) simplifies the form of observables of interest.

The Schr\"{o}dinger equation
    \begin{equation}\label{eq:schrodinger}
    i\hbar \partial_{t} \bra{\Psi(t)} = \hat{\mathcal{H}}\bra{\Psi(t)}
    \end{equation}
cannot be solved analytically. We solve it numerically in the Fock state basis with fixed number of particles $N$. 
A pure state can be decomposed in the Fock state basis
$\bra{\Psi(t)} = \sum\limits_{k=0}^{N}c_{k}(t)\bra{k,N-k}$, 
and then coupled first-order differential equations for the coefficients $c_{k}$ can be solved using matrix exponential. 
In this way, one can forget about an abstract Hilbert space and consider only the representation of operators and states in a more familiar vector space. 
Operations such as dot product, addition and multiplication transfer into the vector space.
Observables are represented as square hermitian matrices, and ket states as column vectors.

The initial state for the evolution is a spin coherent state written in the Fock basis \cite{Diaz2012},
    \begin{align}
    \bra{\theta, \varphi} = \sum\limits_{k=0}^{N}\binom{N}{k}^{1/2}\left[\cos\left(\frac{\theta}{2} \right)\right]^{k}&\left[\sin\left( \frac{\theta}{2}\right)e^{i\varphi} \right]^{N-k}  \nonumber\\[3mm]
    & \times \bra{k,N-k}, \label{eq:coherent}
    \end{align}
and is parametrized by the two real variables $0 \leq \theta < \pi$ and $-\pi \leq \varphi < \pi$.

In what follows, we will concentrate on two physical quantities: (i) the spin-squeezing parameter and (ii) the quantum Fisher information.

The spin squeezing parameter, for mixed and pure states, is defined as~\cite{Wineland1992}
\begin{equation}
\xi^{2}=\frac{N \var{\hat{S}_{\perp}}_{\text{min}}}{|\me{\hat{\vec{S}}}|^{2}}
\label{eq:xiR}
\end{equation}
where $N$ is the total atom number and $\var{\hat{S}_{\perp}}_{\text{min}}$ is the minimal variance of the spin component normal to the mean spin vector $\me{\hat{\vec{S}}}$. 
The state is referred as the spin squeezed state when $\xi^2<1$.

The quantum Fisher information is an important quantity in interferometry. In general, the output state $\hat{\rho}_{\text{out}}$ of an interferometer is
    \begin{equation}
    \hat{\rho}_{\text{out}} = e^{-i \theta \hat{S}_{\vec{n}}} \hat{\rho}_{\text{in}} e^{i \theta \hat{S}_{\vec{n}}}\text{,}
    \end{equation}
where $\hat{S}_{\vec{n}} = \hat{\vec{S}} \cdot \vec{n}$ and $\vec{n}$ is a unit vector representing the effective rotation axis in a given interferometric sequence. 
The precision of the phase shift $\Delta \theta$ depends on the input state $\hat{\rho}_{\text{in}}$, chosen estimator and measurement performed on the output state \cite{Ferrini2011}. 
According to the Cram\'{e}r-Rao inequality, there is a lower bound on the precision with which the phase shift can be determined
$\Delta \theta \geq 1/\sqrt{m F_{_Q}}$,
where $m$ is the number of measurements and $F_{_Q}$ is the quantum Fisher information.
The quantum Fisher information, for a pure state, is given by~\cite{Ferrini2011}
\begin{equation}
F_{_Q}= 4 \var{\hat{S}_{\vec{n}}}_{\text{max}},
\end{equation}
with $\var{\hat{S}_{\vec{n}}}_{\text{max}}$ being the maximal variance of the spin component optimized over all possible directions $\vec{n}$.
%If the quantum Fisher information is larger than the total particle number then the state is entangled \cite{Pezze2014, Ferrini2011}.
The quantum Fisher information is equal to $F_{_Q} = N$ for the spin coherent state, and corresponds to the shot-noise limit of the phase estimation precision in optical interferometry or to the projection noise in the atomic equivalent. The highest possible precision one can achieve is the Heisenberg limit $\Delta \theta \geq 1/\sqrt{m}N$, with $F_{_Q} = N^{2}$. 
This limit can be reached using only a highly entangled state, for example the N00N state.

Both quantities we are interested in are linked to the multiparticle entanglement. If the spin squeezing parameter is smaller than unity, or if the quantum Fisher information is larger than the particle number, then the state of the system is entangled \cite{PezzeSmerzi, Giovannetti, Ferrini2011, Pezze2014}. Moreover, the quantum Fisher information recognizes all entangled states which are useful for high-sensitivity interferometry.

Before we proceed to analyze the quantum dynamics, let us focus on the mean-field approximation. 
Although quantum fluctuations are lost in this description, we argue that the knowledge of the classical phase space dynamics provides an invaluable tool for the detection of useful states in quantum metrology.

% -------------------------------------------
% Section 3: mean-field approximation
% -------------------------------------------
\section{Mean-field phase space}

The mean-field phase space dynamics is a good navigator for the dynamical spin squeezing. It was shown that quantum evolution distinguishes between stable and unstable classical fixed-points \cite{Shchesnovich2007, Shchesnovich2008}, and quantum Hamiltonian eigenstates localize on classical phase space energy contours \cite{Trimborn2009}. 
%Furthermore, two basic mechanisms of squeezing namely one-axis twisting and two-axis counter-twisting correspond to unstable mean-field fixed points. 
%It does not mean that squeezing can not occur in classical stable fixed points, as we emphasize later.

In the limit of large system size $N\gg 1$, we replace bosonic creation and annihilation operators by \textit{c}-numbers \cite{Smerzi_1997}
    \begin{equation}
    \hat{a} \rightarrow \sqrt{N}\sqrt{\rho_{a}}e^{i\varphi_{a}}\text{,}\ \ \hat{b} \rightarrow \sqrt{N}\sqrt{\rho_{b}}e^{i\varphi_{b}}\, .
    \end{equation}
The fixed number of particles $N = a^{\dagger}a + b^{\dagger}b$, dictates the normalization condition $\rho_{a} + \rho_{b} = 1$. Two canonical variables, the population difference $z = \rho_{a} - \rho_{b}$ and the relative phase $\varphi = \varphi_{b} - \varphi_{a}$, are sufficient to describe classical dynamics \cite{Raghavan}. Notice, when one of the modes is fully populated ($z = 1$ or $z = -1$), the relative phase is not well defined. Spin operators become
    \begin{subequations}
        \begin{align}
        \hat{S}_{x} &\rightarrow \frac{N}{2}\sqrt{1 - z^{2}}\cos\varphi,\\
        \hat{S}_{y} &\rightarrow \frac{N}{2}\sqrt{1 - z^{2}}\sin\varphi,\\
        \hat{S}_{z} &\rightarrow \frac{N}{2}z,
        \end{align}
    \end{subequations}
with the mean-field Hamiltonian
    \begin{equation}\label{eq:mean_field}
    H = \me{\hat{\mathcal{H}}} \rightarrow - \hbar \chi \frac{N^{2}}{2} z\sqrt{1-z^2}\sin\varphi.
    \end{equation}

Equations of motion for the canonical position $\varphi$ and the conjugate momentum $z$ can be derived from quantum mechanical Heisenberg equations or classical Hamilton equations,
    \begin{subequations}
    \begin{align}
    \frac{d \varphi}{dt} &= \frac{2}{\hbar N}\frac{\partial H}{\partial z} = -N\chi\frac{1-2z^2}{\sqrt{1-z^2}}\sin\varphi\, ,\\
    \frac{d z}{dt} &= -\frac{2}{\hbar N}\frac{\partial H}{\partial \varphi} = N\chi z\sqrt{1-z^2}\cos\varphi\, .
    \end{align}
    \label{eq:motion}
    \end{subequations}
\noindent Instead of solving these coupled differential equations we will analyze the topology of the phase portrait. The phase portrait is just a geometrical representation of trajectories of a dynamical system in the phase space. In our case, trajectories are tangent to the velocity field $(\dot{\varphi}, \dot{z})$.% because decoherence is not included in the formalism.

  \begin{figure}[]
    \centering
    \includegraphics[width = 0.49\textwidth]{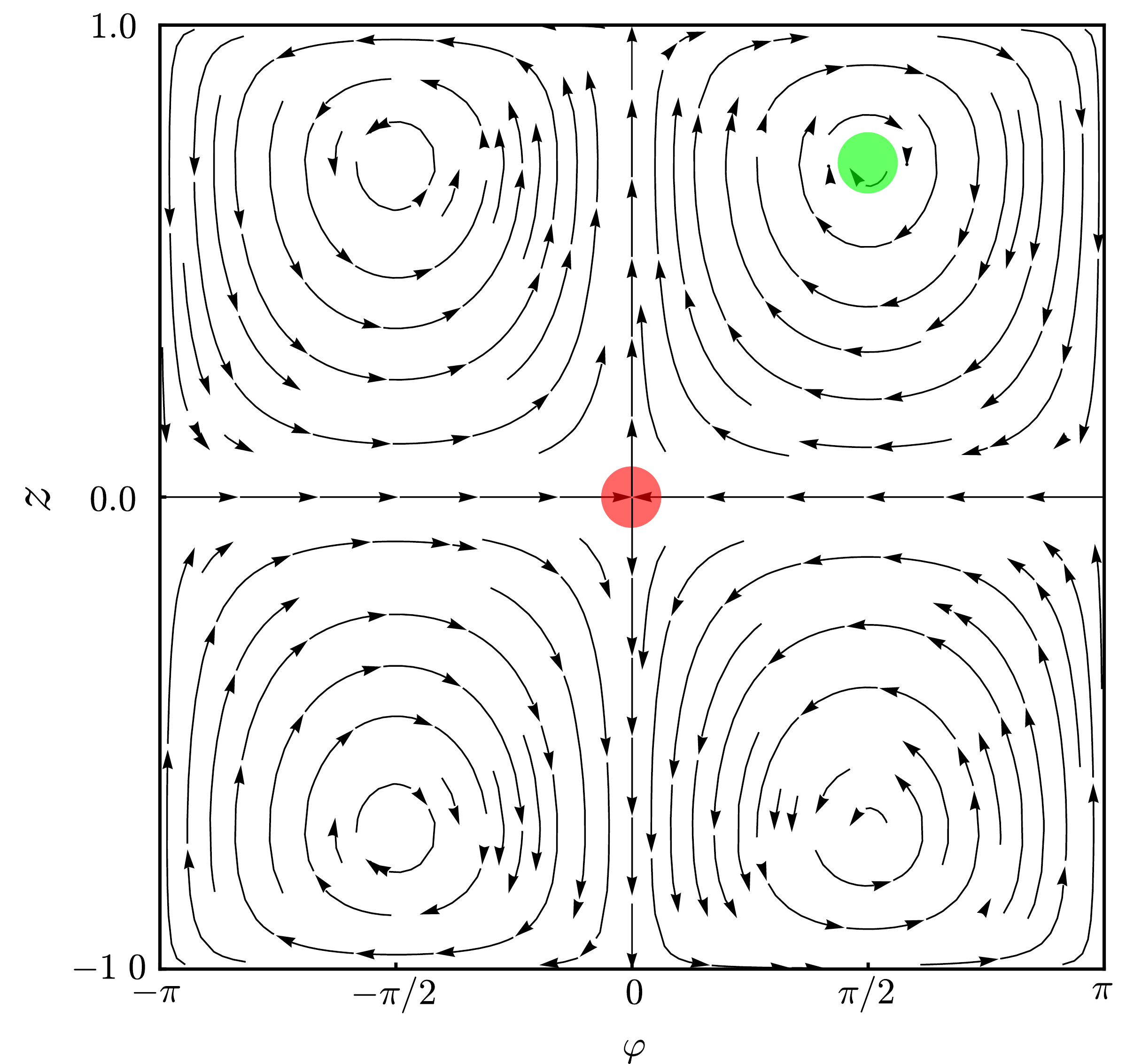}
    \caption{Mean-field trajectories of the two-axis counter-twisting Hamiltonian $\hat{\mathcal{H}} = -\hbar\chi(\hat{S}_{z}\hat{S}_{y} + \hat{S}_{y}\hat{S}_{z})$. Two unstable saddle fixed points are located at $z = 0$ and $\varphi = 0,\pi$. Four neutrally stable center fixed points correspond to $(\varphi,z) = (\pm\pi/2, \pm 1/\sqrt{2})$. 
Colored spots visualize the spin coherent states around the classical unstable (red) and stable (green) fixed points.}
    \label{fig:two_axis_phase_portrait}
  \end{figure}

The phase portrait in the two-dimensional system consists of fixed points or closed orbits. Fixed or equilibrium points correspond to a steady state, and satisfy
    \begin{equation}
        \begin{cases}
            \frac{1-2z^2}{\sqrt{1-z^2}}\sin\varphi = 0\, ,\\[4mm]
            z\sqrt{1-z^2}\cos\varphi = 0\, .
        \end{cases}
    \end{equation}
Close to a stationary point, equations of motion (\ref{eq:motion}) can be linearized and solved exactly. Information about stability of fixed points can be determined from a stability matrix \cite{Xiao2012}. Depending on eigenvalues and eigenvectors of the stability matrix one can classify fixed points according to the behavior of nearby trajectories.
The phase portrait for (\ref{eq:mean_field}) presented in Fig.~\ref{fig:two_axis_phase_portrait} consists of two unstable saddle fixed points located at $z = 0$ and $\varphi = 0,\pi$, and four stable center fixed points at $(\varphi,z) = (\pm\pi/2, \pm 1/\sqrt{2})$. 

%Quantum dynamics and squeezing depend strongly on the location of initial coherent spin state. Their location at unstable fixed points give strong squeezing and wealth of quantum states for high-precision atom interferometry while at stable fixed points give quantum states that are useless for these purposes.

% ----------------------------------------
% section 4: unstable fixed point
% ----------------------------------------
\section{Dynamics around an unstable fixed point}

We start with the spin coherent state located at an unstable saddle fixed point on the equator, $\bra{\pi/2,0} = e^{-i\frac{\pi}{2}\hat{S}_{y}}\bra{N,0} = \bra{N,0}_{x}$, being the eigenstate of the $\hat{S}_{x}$ operator with eigenvalue $N/2$. Location of the initial state in corresponding mean-field phase space is sketched in Fig. \ref{fig:two_axis_phase_portrait}.
The red spot represents the initial spin coherent state, black lines are mean-field trajectories of the Hamiltonian, while the arrows indicate the direction of the evolution. 
The angle between inflowing and outflowing trajectories in the phase space is $\pi/2$.
The regular dynamics takes place for short times.
The initial state is stretched along the meridian of the Bloch sphere leading to the highly reduced variance of the $\hat{S}_y$ component of the spin operator and highly increased variance of the $\hat{S}_z$ component. The best squeezing occurs at this stage of the evolution, and useful states for high-precision measurements are generated. 
Moreover, the squeezing parameter is determined by the variance of the $\hat{S}_y$ component of the spin operator, while the quantum Fisher information is determined by the variance of the $\hat{S}_z$ component. 
Later, the variances of the $\hat{S}_y$ and $\hat{S}_z$ components of the spin become once again of the same order with $\me{\hat{S}_x} \simeq-N/2$. 
Next, stretching of the state takes place but the direction of the evolution is opposite. 
The dynamics becomes irregular, and results depend on the total number of particles.
%Husimi and Wigner maps for characteristic times of evolution are presented in fig.\ref{fig:husimi_wigner_two_axis}.

%\begin{figure}[]
%\centering
%\includegraphics[width = 0.3\textwidth]{fig2.pdf}
%\caption{Visualization of the initial spin coherent state (red spot) on the Bloch sphere around the classical unstable fixed point.
%Mean-field trajectories are marked by gray lines while arrows indicate the direction of the evolution.}
%\label{fig:two_axis1}
%\end{figure}

%-----------------------------------------------------------
\subsection{Scaling with the system size}
\label{subsec:scaling}

In order to analyze scaling of the spin squeezing parameter and the quantum Fisher information with the system size we use a general theory developed in \cite{AnglinVardi, AndreLukin}.
One starts with equations of motion for operators of spin components $\me{ \dot{\hat{S}}_j  }$ which involve terms that depend on the first-order moment 
$\me{\hat{S}_j}$ and second-order moments $\me{\hat{S}_j \hat{S}_k}$. Then, the time evolution of the second-order moments depends on themselves and on third-order moments, and so on. 
It leads to the Bogoliubov-Born-Green-Kirkwood-Yvon (BBGKY) hierarchy of equations of motion for expectation values of operator products. 
We truncate the hierarchy by keeping the first- and the second-order moments,
\begin{eqnarray}
\me{\hat{S}_i \hat{S}_j \hat{S}_k} &\simeq & \me{\hat{S}_i \hat{S}_j} \me{\hat{S}_k} + \me{\hat{S}_j \hat{S}_k} \me{\hat{S}_i} + \me{\hat{S}_k \hat{S}_i} \me{\hat{S}_j}  \nonumber \\
{}&-&2 \me{\hat{S}_i} \me{\hat{S}_j}\me{\hat{S}_k}.
\end{eqnarray}

Let us first introduce a small parameter $\varepsilon=1/N$ and transform spin components into $\hat{h}_j=\sqrt{\varepsilon} \hat{S}_j$; then the Hamiltonian reads
\begin{equation}
\hat{H}=\frac{\chi}{\varepsilon} \left( \hat{h}_y \hat{h}_z + \hat{h}_z \hat{h}_y \right),
\end{equation}
and commutation relations are $[\hat{h}_i, \hat{h}_j]=i \varepsilon \hat{h}_k \epsilon_{ijk}$.

Equations of motion for expectation values $s_j=\me{\hat{h}_j}$ and second order moments 
$\delta_{jk}=\me{\hat{h}_j \hat{h}_k + \hat{h}_k \hat{h}_j}- 2 \me{\hat{h}_j}\me{\hat{h}_k}$ relevant for our purposes are
\begin{subequations}
\begin{align}
\dot{s}_x&=2\left( \delta_{yy} - \delta_{zz} \right) ,\label{eq:sx} \\
\dot{\delta}_{yy}&=-4\, \delta_{yy} \,  s_x , \label{eq:dyy}\\
\dot{\delta}_{zz}&=4\, \delta_{zz} \, s_x , \label{eq:dzz}
\end{align}
%\label{eq:motion}
\end{subequations}
where we have introduced the dimensionless time $\tau=\chi t/\sqrt{\varepsilon}$.
The initial coherent state at the unstable saddle fixed point, $\bra{\pi/2,0}$, gives the following initial conditions: $s_x(0)=1/2\sqrt{\varepsilon}$ and $\delta_{yy}(0)=\delta_{zz}(0)=1/4$.
Equation (\ref{eq:sx}) takes the form $\dot{s}_x(\tau)=-\sinh \left[ f\left( \tau \right) \right]$ with $f\left( \tau \right)=4 \int_0^{\tau} s_x(\tau') d\tau'$, 
that has an analytical solution when one expands the function $f$ up to the first order in Taylor series, $f(\tau)\simeq f(0) + f'(0)\tau$. 
This requires $\tau\ll 1$ or $\chi t \ll 1/\sqrt{N}$. The self-consistency condition gives $f(0)=0$ and $f'(0)=4 s_x(0)$, and the approximated solution for $s_x$ takes the form
\begin{equation}
s_x(\tau)=s_x(0) - \frac{\cosh \left[ 4 s_x(0) \tau \right]-1}{4 s_x(0)} \,. 
\label{eq:sxsol}
\end{equation}
The above expression together with equations (\ref{eq:dyy}) and (\ref{eq:dzz}) gives the squeezing parameter
\begin{equation}
\xi^2 = e^{-4 s_x(0)\tau + \left( \sinh \left[ 4 s_x(0)\tau \right] - \tau \right)/4 s_x(0)^2} \, ,
\label{eq:xisol}
\end{equation}
and the time-dependent quantum Fisher information
\begin{equation}
F_{_Q} (\tau)= \frac{1}{\varepsilon} e^{4 s_x(0)\tau - \left( \sinh \left[ 4 s_x(0)\tau \right]-\tau \right)/4 s_x(0)^2} \, .
\label{eq:FQsol}
\end{equation}
Notice, two approximations were made to obtain solutions (\ref{eq:sxsol})-(\ref{eq:FQsol}). 
The first is the truncation of the BBGKY hierarchy, which is equivalent to the Gaussian approximation, 
and the second is the short-time expansion which limits the validity of the solution to $\tau\ll 1$.

Minimization of the squeezing parameter (\ref{eq:xisol}) over the time gives scaling of the best squeezing time 
$\tau_{{\rm best}}\simeq \ln \left(8 s_x(0)^2 \right)/4 s_x(0)$ or $\chi t_{{\rm best}}\simeq \ln(2N)/2N$. The same holds for the first maximum of the quantum Fisher information.
Expansion of the squeezing parameter at the best squeezing time $\tau_{{\rm best}}$ in terms of the small parameter $\varepsilon$ gives
\begin{equation}
\xi^2_{\rm best}\simeq \frac{e}{2} \varepsilon ,
\end{equation}
which reproduces the known result $\xi^2_{\rm best}\propto N^{-1}$. Leading terms of the maximum of the quantum Fisher information at the best time are
\begin{equation}
F_{_Q, \, {\rm best}} \simeq \frac{2}{e} \frac{1}{\varepsilon^2} ,
\end{equation}
and provide the Heisenberg-like scaling $F_{_Q, \, {\rm best}} \simeq 0.73 N^2$.

\begin{figure}[]
\centering
\includegraphics[width = 0.45\textwidth]{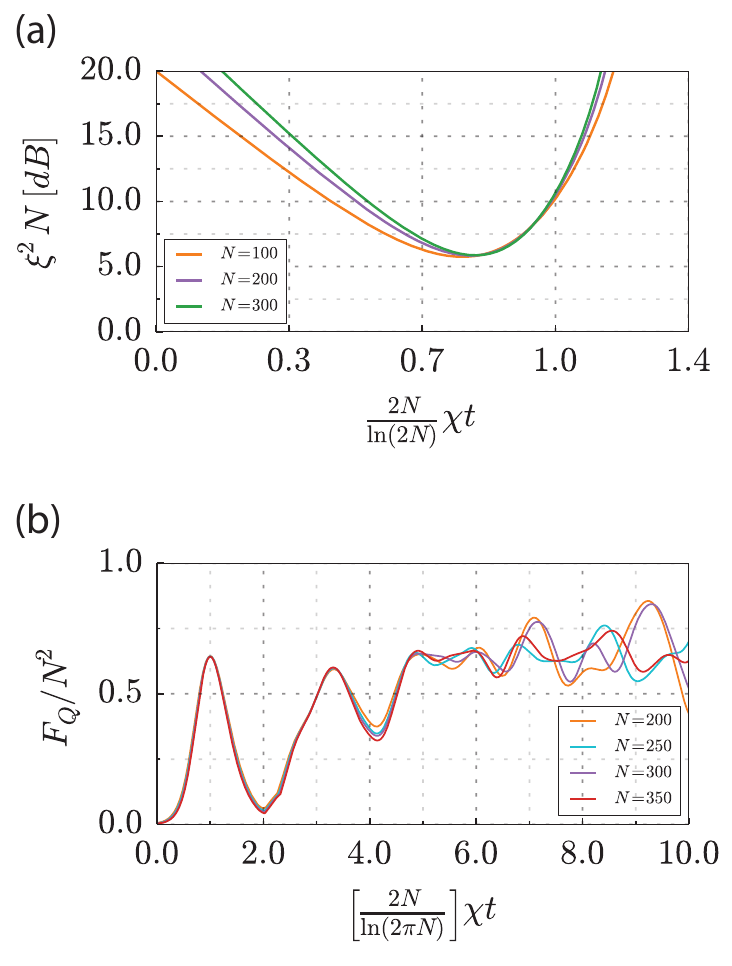}
\caption{Scaling of the squeezing parameter (a)
and the quantum Fisher information (b) with the total number of particles $N$.}
\label{fig:squeezing_two_axis}
\end{figure}

In Fig. \ref{fig:squeezing_two_axis} we show the squeezing parameter and the quantum Fisher information for different number of particles $N$ in the rescaled time.
We have shown only a small period of time where scaling with the number of particles may agree with our simulations. For further moments there is no time regularity. 

The scaling prediction for the squeezing parameter is surprisingly correct, the minima of $\xi^2$ nicely coincide for different particle numbers. 
The prediction of the model for the time scaling of the first maximum of the quantum Fisher information is not perfect. 
First of all, the simple model predicts the position of the first maximum of the quantum Fisher information at the same time as the minimum of the squeezing parameter, 
since we know from the exact numerical calculations that the maximum is located at later times. Nevertheless, the time scaling of type 
$\chi t_{\rm best} \sim \ln(aN)/(bN)$ survives, and we find values of $a=2\pi$ and $b=2$ giving the best result. 
The first maximum of the quantum Fisher information is achieved at $F_{_Q, {\rm best}}\sim 0.67 N^2$ in numerical simulations, the width of the peak decreases
with larger number of particles. 

In addition, the level of squeezing achieved by the two-axis counter-twisting model is always better than the squeezing generated by the one-axis twisting Hamiltonian, the last scales like $\sim N^{-2/3}$.
The first maximum of the quantum Fisher information given by the two-axis counter-twisting model appears before the characteristic plateau of the quantum Fisher information generated by the one-axis twisting Hamiltonian which occurs at $\chi t\sim 1/\sqrt{N}$.

%-------------------------------------------------------------
\subsection{Quantum states for high-precision measurements}

\begin{figure}[]
\centering
\includegraphics[width = 0.45\textwidth]{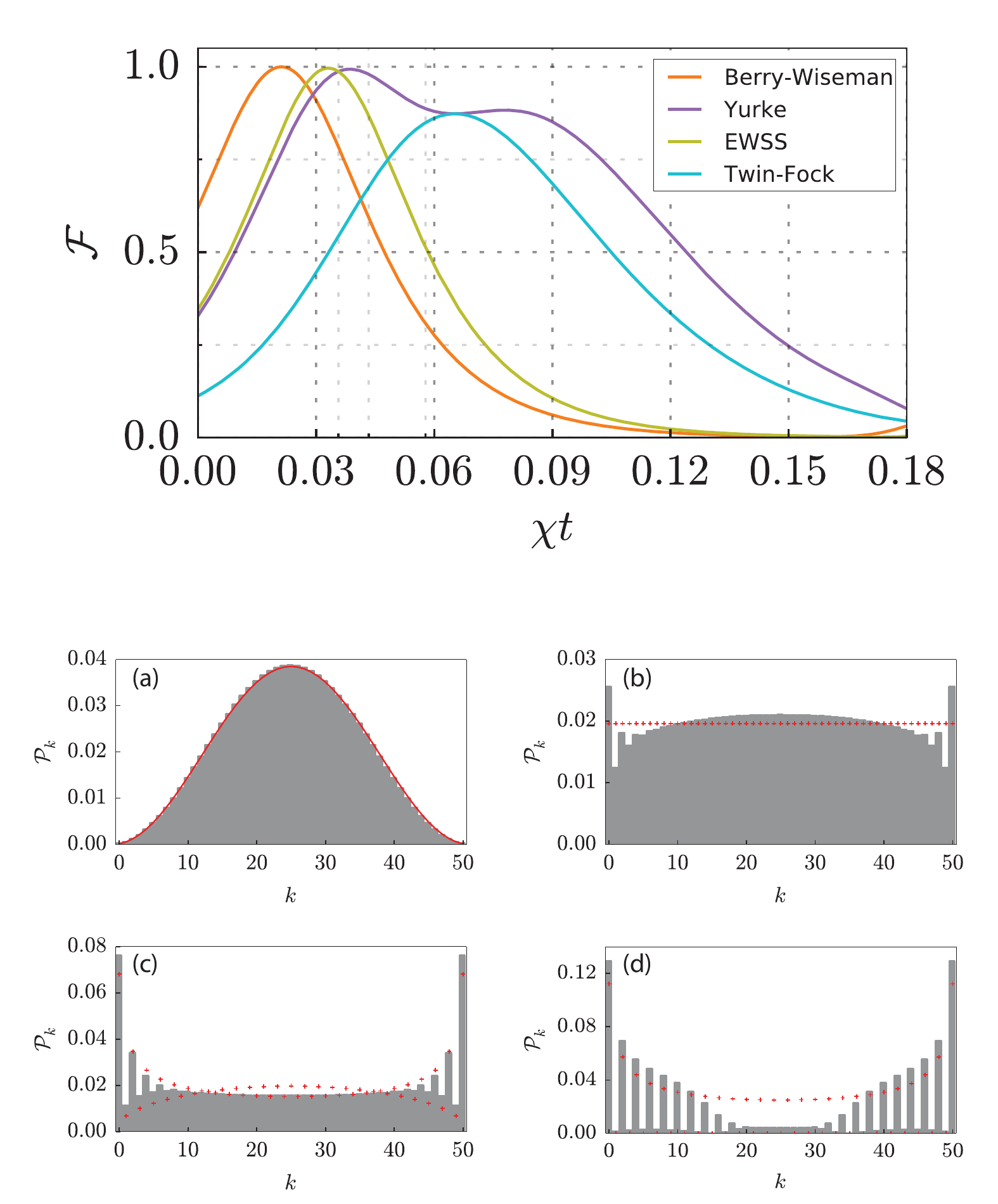}
\caption{Top panel: Time evolution of the fidelity (\ref{eq:fidelity}) to the EWSS, Twin-Fock, Berry-Wiseman and optimized over $\alpha$ Yurke states. The maxima of respective fidelity functions are: $\mathcal{F}_{\rm BW,\, max} = 0.9999$ for the BW state, $\mathcal{F}_{\rm EWSS,\, max} = 0.9965$ for the EWSS state, 
$\mathcal{F}_{\rm Y,\, max} = 0.9936$ for the Yurke state with $\alpha = 0.678$, and $\mathcal{F}_{\rm TF,\, max} = 0.8732$ for the twin-Fock state.
Bottom panels: Probability distributions (\ref{eq:probability}) with Fock states at the maximum of the respective fidelity function for (a) Berry-Wiseman, (b) EWSS, (c) Yurke and (d) Twin-Fock states. Red crosses correspond to the ideal case given by equations (\ref{eq:BW})-(\ref{eq:TF}). The total number of particles is $N = 50$.}
\label{fig:comp_fidelity}
\end{figure}

States produced by the two-axis counter-twisting Hamiltonian, even if they may not have a simple analytical form, are particularly useful for high-precision measurements since they provide Heisenberg-like scaling of the phase sensitivity $\Delta \theta$. Here and below, we list a few particular states, expanded in the Fock state basis, that are generated by the TACT Hamiltonian:
\begin{itemize}
\item[(i)] the Berry-Wiseman state (BW)~\cite{Berry2000}
\begin{equation}
\bra{{\rm BW}} = \frac{1}{\sqrt{1+N/2}}\sum\limits_{k=0}^{N}\cos\left[\frac{(k-N/2)\pi}{N+2} \right]\bra{k,N-k}\text{,}
\label{eq:BW}
\end{equation}
that gives the quantum Fisher information 
\begin{equation}
F_{_Q}=\frac{2}{2+N}\sum_{k=0}^N \cos^2\left[\frac{(k-N/2)\pi}{N+2} \right]\left(2k -N \right)^2,\nonumber
\end{equation}
which for $N\gg 1$ is $F_{_Q}\approx0.13 N^2$,
\item[(ii)] the equally weighted superposition state (EWSS)~\cite{Yukawa2014}
\begin{equation}
\bra{{\rm EWSS}} = \frac{1}{\sqrt{N+1}}\sum\limits_{k=0}^{N}\bra{k,N-k}
\label{eq:EWSS}
\end{equation}
with $F_{_Q} = \left( N^{2}/3 \right) \left(1 + 2/N \right)\approx0.33N^{2}$,
\item[(iii)] the Yurke state (Y) for even number of particles~\cite{Combes2005}
\begin{eqnarray}
\bra{{\rm Y}} &=& \frac{\sin\alpha}{\sqrt{2}}\bra{N/2+1, N/2-1} + \cos\alpha \bra{N/2,N/2} \nonumber \\
{}&+&\frac{\sin\alpha}{\sqrt{2}}\bra{N/2-1, N/2+1},
\label{eq:Y}
\end{eqnarray}
with some real parameter $\alpha$ and the quantum Fisher information of the form
$F_Q=\left( N/2 \right) \left( \left( N/2+1 \right) \left( 2-\sin^2\alpha \right)-2\sin^2\alpha \right)$,
\item[(iv)] the twin-Fock state (TF) for even number of particles~\cite{Yukawa2014}
\begin{equation}
\bra{{\rm TF}}= \bra{N/2, N/2}
\label{eq:TF}
\end{equation}
with $F_{_Q} = \left( N^{2}/2 \right) \left(1 + 2/N \right) \approx 0.5N^{2}$.
\end{itemize}
In order to demonstrate their appearance during the evolution, in Fig.~\ref{fig:comp_fidelity} we show fidelity functions defined as
\begin{equation}
\mathcal{F}_{{\rm A}}(t) = |\langle\left. {\rm A} \right|e^{-i\hat{\mathcal{H}} t}\left| N,0\right\rangle_{x}|^{2}\text{,}
\label{eq:fidelity}
\end{equation}
where ${\rm A}$ is one of the states of interest (\ref{eq:BW})-(\ref{eq:TF}). The Twin-Fock and Yurke states were rotated with $\hat U = e^{-i \pi \hat{S}_x/2 }$ to maximize $\mathcal{F}$. In addition, in Figs.~\ref{fig:comp_fidelity}a-\ref{fig:comp_fidelity}d we also show 
the probability distribution with Fock states
\begin{equation}
\mathcal{P}_{k} = |\langle\left. k,N-k\right|\left. \Psi \right\rangle|^{2}\text{,}
\label{eq:probability}
\end{equation}
calculated at the maximum of respective fidelity functions (in gray) compared to the exact probability of the respective state (in red).

\begin{figure*}[]
\centering
\includegraphics[width = 0.65\textwidth]{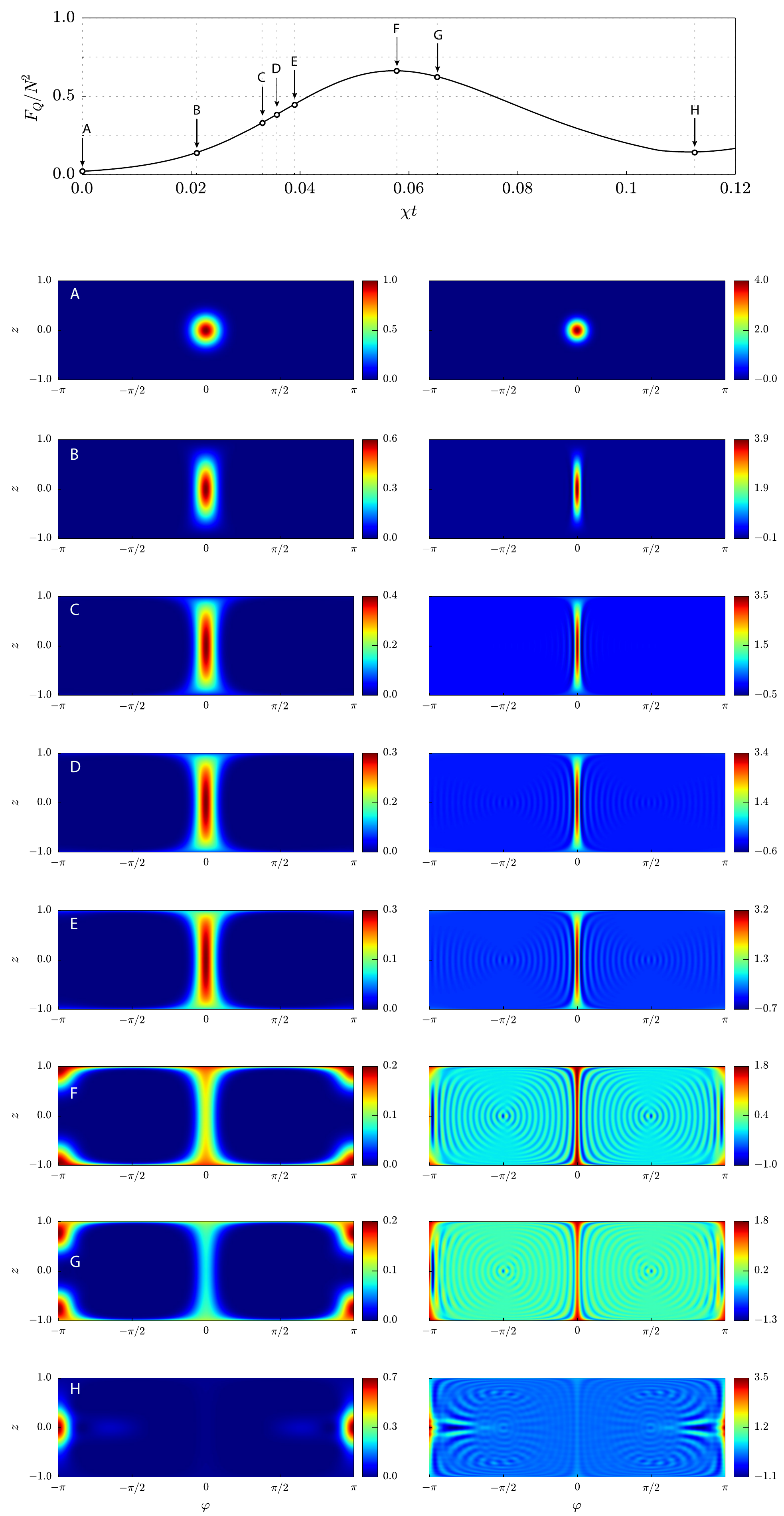}
\caption{
Top panel: arrows indicate the locations of quantum states that are plotted in bottom panels on the time evolution of the quantum Fisher information. 
Bottom panels: the Husimi (left column) and Wigner (right column) functions at particular moments of time:
(A) the initial coherent spin state, 
(B) the maximum of the BW fidelity function, 
(C) the maximum of the EWSS fidelity function, 
(D) the best spin squeezing, 
(E) the maximum of the Yurke fidelity function, 
(F) the maximum of the quantum Fisher information, (G) the maximum of the twin-Fock fidelity function, and (H) the second local minimum of the quantum Fisher information. The total number of particles is $N = 50$.}
\label{fig:husimi_wigner_two_axis}
\end{figure*}

The two-axis counter-twisting Hamiltonian produces almost ideal Berry-Wiseman, EWSS and Yurke states, since the maximum of the fidelity function is one, see the top panel in Fig. \ref{fig:comp_fidelity}. Nevertheless, the probability distribution with Fock states is exact only for the BW state \cite{Combes2005, Jin2010}, while for the EWSS and Yurke states very small differences can be found. The maximum of the fidelity to the twin-Fock state is $\mathcal{F}_{\rm TF,\, max} = 0.8732$, showing that the generated state is not the perfect TF state. 
The Yurke state becomes the twin-Fock state for $\alpha=0$ what explains their equal fidelities $\mathcal{F}$ at $\chi t=0.065$.
Notice, the maxima of the fidelity functions for the BW, EWSS and Yurke states are reached at times very close to the best squeezing times (just before the first maximum of the quantum Fisher information). The maximum of the fidelity function for the TF state is located at times larger than the first maximum of the quantum Fisher information. 
The maxima of fidelity functions of particular states support the scaling of type $\sim \ln(aN)/(bN)$, where the coefficients $a,b$ may depend on the state~\cite{Yukawa2014}. Another interesting observation is near the unity value of the fidelity function for the N00N state (not shown in figures). Position of the  maximum is not regular and shifts in time as the number of particles changes.

In order to get a better insight into the regular part of the quantum dynamics we plot the Husimi distribution function~\cite{Husimi1940}
   \begin{equation}
   Q(t,\theta,\phi) = |\langle\left. \theta,\phi \right|\left.\Psi(t) \right\rangle|^{2},
   \end{equation}
and the Wigner function $W(\theta,\phi)$~\cite{Agarwal1994} at different moments of time. 
The top panel in Fig. \ref{fig:husimi_wigner_two_axis} shows a location of particular states on the time evolution of the quantum Fisher information, while the bottom panels present their Husimi (in the left column) and Wigner (in the right column) function maps. The Husimi distribution function is unique, regular and positive definite. Moreover, the position and number of zeros of the Husimi function contains all information about a pure quantum state \cite{Gagen, Zyczkowski}. This structure is visible in the logarithmic scale only under the quite high precision of calculations, and is invisible in Fig.~\ref{fig:husimi_wigner_two_axis}. The Wigner function of a quantum state, in addition to localized maxima, may have interference fringes of negative value. 
We observe increasing formation of the interference fringes during quantum evolution, even if their details may not be visible on a linear scale in Fig.~\ref{fig:husimi_wigner_two_axis}.
Number of fringes is equal to the total particle number $N$. Moreover, the position of the maxima is the same starting from EWSS till TF states. It is the amplitude of the interference fringes that 
distinguish particular states. Negativity of the Wigner function was shown to be an indicator of non-classicality of a quantum state \cite{Zyczkowski2}. Thus, it is a convenient tool for detection of useful states for precision measurements.

The Wigner function can be used to justify the large value of the quantum Fisher information. Typically, the presence of oriented fringes in the Wigner map increases its value.
It will be clear when we will recover that the Fisher information has its geometrical interpretation \cite{Braunstein1994, Ferrini2011, FerriniPhd2011}. If one takes
    \begin{equation}
    \hat{\rho}_{\text{out}} = e^{-i d\theta \hat{S}_{\vec{n}}} \hat{\rho}_{\text{in}} e^{i d\theta \hat{S}_{\vec{n}}}
    \end{equation}
with infinitesimal angle $d\theta$ then the Bures-Riemannian metric becomes
    \begin{equation}
    d^{2}(\hat{\rho}_{\text{in}}, \hat{\rho}_{\text{out}}) = F_{Q}\left[\hat{\rho}_{\text{in}}, \hat{S}_{\vec{n}} \right]d\theta^{2}\text{.}
    \end{equation}
The larger the quantum Fisher information, the faster the state $\hat{\rho}_{\text{out}}$ becomes distinguishable from $\hat{\rho}_{\rm in}$. Because the Wigner function is a bijective mapping we can capture sensitivity to rotations of a given quantum state.
The Wigner function graphically justifies which measurement $S_{\vec{n}}$ optimizes the quantum Fisher information.
If we look at the Wigner function of some state, it will be apparent that the rotation of the state through an infinitesimal angle $d\theta$ around $\hat{S}_{\vec{n}}$ shifts minima and maxima of interference fringes making the Bures distance large.

% ----------------------------------------
% section 5: stable fixed point
% ----------------------------------------

\section{Dynamics around a stable fixed point}

%  \begin{figure}[]
%    \centering
%    \includegraphics[width = 0.3\textwidth]{fig6.pdf}
%    \caption{Visualization of the initial spin coherent state (red spot) on the Bloch sphere around the classical stable fixed point for the Hamiltonian $\hat{\mathcal{H}} = -\hbar\chi(\hat{S}_{z}\hat{S}_{y} + \hat{S}_{y}\hat{S}_{z})$. Equal energy regions are marked by gray lines while arrows indicate the direction of evolution. The mean spin vector remains almost constant during the evolution time.}
%    \label{fig:two_axis2}
%  \end{figure}

The initial spin coherent state located at the classical stable fixed point is visualize in Fig.~\ref{fig:two_axis_phase_portrait} by green spot.
The state is slightly deformed in time leading to the Husimi function that changes regularly between the circular and elliptic shape. 
The frequency can be easily estimated within the frozen spin approximation \cite{SongLi2011, Yiang2013}.
In this case it is more convenient to analyze an equivalent to (\ref{eq:two_axis_hamiltonian_exact}) Hamiltonian, namely,
  \begin{equation}
    \hat{\tilde{\mathcal{H}}} = \hbar\chi\left(\hat{S}_{x}^{2} - \hat{S}_{y}^{2} \right)\text{,}
  \end{equation}
because stable fixed points are located on the equator and calculations simplify significantly, 
$\hat{\tilde{\mathcal{H}}} = \hat{U} \hat{\mathcal{H}}\hat{U}^{\dagger}$ with $\hat{U} = e^{-i (\pi/2) \hat{S}_{y}} e^{-i (\pi/4) \hat{S}_{x}}$.

\begin{figure}[t]
\centering
\includegraphics[width = 0.45\textwidth]{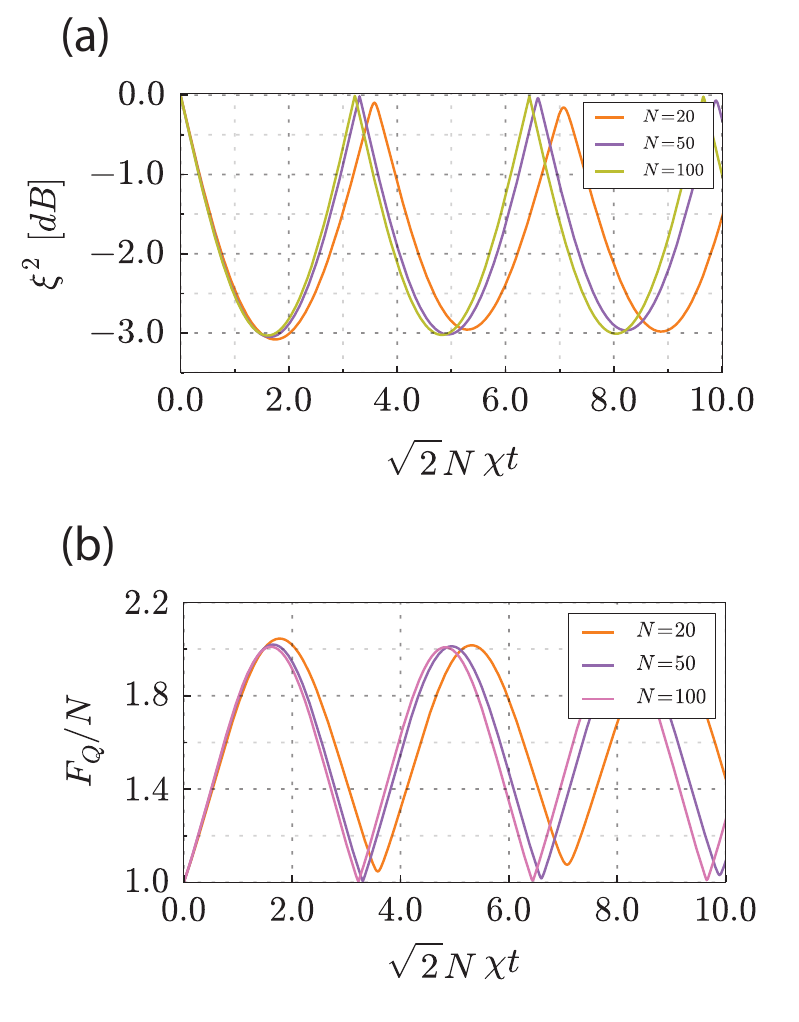}
\caption{The spin squeezing parameter (a) and the quantum Fisher information (b) for different numbers of particles $N$ around a stable fixed point.}
\label{fig:stable}
\end{figure}

In the frozen spin approximation one starts with equations of motion for spin operators
    \begin{subequations}
        \begin{align}
        \frac{d}{dt} \hat{S}_{x} &= \chi \left(\hat{S}_{y}\hat{S}_{z} + \hat{S}_{z}\hat{S}_{y}\right)\text{,}\\
        \frac{d}{dt}\hat{S}_{y} &= -\chi \left( \hat{S}_{x}\hat{S}_{z} + \hat{S}_{z}\hat{S}_{x}\right)\text{,}\\
        \frac{d}{dt}\hat{S}_{z} &= 2\chi\left(\hat{S}_{x}\hat{S}_{y} + \hat{S}_{y}\hat{S}_{x} \right)\text{.}
        \end{align}
    \end{subequations}
The initial condition $\bra{\pi/2,0}$ corresponds to $\me{\hat{S}_{x}} = N/2$ and the dynamics will be captured around this point. 
In the frozen spin approximation one replaces the operator $\hat{S}_{x}$ by its mean value $\me{ \hat{S}_{x}} = N/2$, and ends up with equations for the remaining operators
    \begin{subequations}
        \begin{align}
        \frac{d}{dt}\hat{S}_{y} &= -\chi N \hat{S}_{z} ,\\
        \frac{d}{dt}\hat{S}_{z} &= 2\chi N \hat{S}_{y}.
        \end{align}
    \end{subequations}
One can easily solve these coupled differential equations
    \begin{subequations}
        \begin{align}
        \hat{S}_{y}(t) &= \hat{S}_{y}(0)\cos(\omega t) - \frac{1}{\sqrt{2}}\hat{S}_{z}(0)\sin(\omega t)\text{,}\\
        \hat{S}_{z}(t) &= \hat{S}_{z}(0)\cos(\omega t) + \sqrt{2} \hat{S}_{y}(0)\sin(\omega t)\text{,}
        \end{align}
    \end{subequations}
with $\omega = \sqrt{2}N\chi$. The spin squeezing parameter $\xi^{2}$ and the optimized quantum Fisher information are
    \begin{subequations}
        \begin{align}
        \xi^{2} &= \frac{N \var{\hat{S}_{y}}}{|s_{x}|^{2}} = 1 - \frac{1}{2}\sin^{2}(\omega t),\\[3mm]
        F_{Q} &= 4 \var{\hat{S}_{z}} = N\left[1 + \sin^{2}(\omega t) \right]\text{.}
        \end{align}
    \end{subequations}
Results of the frozen spin approximation can be compared to the exact numerical simulations presented in Fig. \ref{fig:stable}. 
Although overall shape of the function slightly deviates from $\sin^{2}(\omega t)$, extreme values and the frequency are in a good agreement.
Now it is clear why the initial spin coherent states located at classical stable fixed points give shot-noise limited sensitivity for quantum metrology. The maximal value of the Fisher information scale linearly with the particle number $N$. It is not possible to beat shot-noise limit in this configuration. 

\section{Summary}

We have discussed the quantum dynamics generated by the two-axis counter-twisting Hamiltonian in the context of quantum metrology.
We have started the analysis within the mean-field description which identifies a convenient location of the initial spin coherent state.
The quantum dynamics around a stable fixed point supports the shot-noise-limited sensitivity for precision measurements. On the other hand, 
the quantum dynamics around an unstable fixed point generates quantum states that give Heisenberg-like scaling. 
We have calculated the spin squeezing parameter and the quantum Fisher information explaining their scaling with the system size.
Our results show that the quantum dynamics with the two-axis counter-twisting Hamiltonian creates quantum correlations in a regular way on a short time scale. 
%The time needed for the creation of useful correlations by the two-axis counter-twisting model is shorter than the time needed by the one-axis twisting Hamiltonian.
In addition, a characteristic feature of the output of TACT Hamiltonian are concentric fringes of negative values in the Wigner functions of states located around the first maximum of the quantum Fisher information. It may be difficult to identify an efficient interferometric strategy and measurement to fully recover this part of the quantum Fisher information. 
In the OAT model, for instance, the quantum Fisher information can be retrieved by a standard measurement of the population imbalance~\cite{Chwedenczuk, Strobel2014}.

\acknowledgments

We thank J. Chwede\'nczuk, K. Gietka, P. Sza\'nkowski, T. Wasak and A. Sinatra, Y. Castin, K. Paw\l{}owski, H. Kurkjian for valuable discussions, and O. Hul for a careful reading of the
manuscript. This work was supported by DEC-2011/03/D/ST2/01938.

% ----------------------------------------
% Bibliography
% ----------------------------------------

\bibliography{bibliografia}
\end{document}